# Detection of contamination in noninvasive prenatal fetal gender test

Gabriel Minárik[1,2,3], Emília Nagyová[3], Gabriela Repiská[2], Tomáz Szemes[1,3], Barbora Vlková-Izrael[1,2], Peter Celec[1,2]

[1] Geneton Ltd., Ilkovičova 3, 841 04 Bratislava, Slovakia

[2] Comenius University in Bratislava Faculty of Medicine, Institute of Molecular Biomedicine, Sasinkova 4, 811 08 Bratislava, Slovakia

[3] Comenius University in Bratislava Faculty of Natural Sciences, Department of Molecular Biology, Mlynska dolina B2-009, 842 15 Bratislava, Slovakia

**Abstract**

The risk of false positive results in noninvasive prenatal diagnosis focused on fetal gender and RhD status determination could be a problem in clinical routine. This is because these tests are based on detection of presence of DNA sequences with high population frequency and so there is the risk of sample contamination during sample collection and processing.

In our study the different fragmentation of fetal and maternal DNA molecules present in maternal circulation was utilized in identification of contaminated samples. Amplification of Y-chromosome specific assays different in size was tested on circulating DNA samples.

Of the four tested assays two shorter (84 and 177 bp) showed expected qPCR efficiency and have comparable amplification profiles. The difference in Ct values between these two assays was found to be statistically significant in comparison of fetal male and normal male samples ($p<0.0001$) as well as in blinded pilot study performed on 10 artificially contaminated and 10 non-contaminated samples ($F=34.4$, $p<0.0001$) that were all identified correctly.

Our results showed that differently sized assays performed well in detection of external contamination of samples in noninvasive prenatal fetal gender test and could be of help in clinical laboratories to minimize the risk of false positive results.

**Introduction**

The detection of cell-free fetal DNA in the maternal circulation during gestation was a milestone [1]. It has revolutionized the field of non-invasive prenatal diagnosis [2]. The original protocol for the detection of fetal DNA was based on the detection of the Y-chromosomal sequences. Although several other approaches are known, this is still the gold standard for a simple detection and quantification of fetal DNA [3].

The main application for the circulating fetal DNA in maternal plasma is the use in the screening for the Down syndrome [4]. This possibility becomes available for many pregnant women with the spread of the high throughput sequencing technology. Even more, it is now possible, although still very costly, to sequence the whole fetal genome from DNA isolated from maternal plasma [5]. However, beyond these methods requiring sophisticated equipment, there are several simple PCR based analyses that can be done.

*RhD* genotyping and fetal sex determination are among those that have already advanced into routine clinical and/or commercial use [6]. The accuracy of the particular analyses reaches 99% and more [7]. However, there is still space for improvement, especially in the determination of fetal sex. It is of commercial interest as many pregnant women want to know the gender of their baby very early just out of curiosity. In several countries this possibility has, of course, ethical aspects that have to be taken into account [8]. The possibility to detect fetal sex early during the first trimester is also of clinical value. It enables a very precise prognosis for the fetus of risk for X-chromosome linked genetic disorders [9].

The non-invasive fetal sex determination is based on a simple PCR targeting the male Y-chromosomal sequences [10]. For the quantification of fetal DNA from a male fetus a single copy gene is used as a target, typically the *SRY* gene. However, multicopy sequences such as *DYS14* are more suitable for qualitative assessments due to higher sensitivity of the PCR. Plasma DNA isolated from non-pregnant women and from pregnant women with a female fetus should not contain any Y-chromosomal sequences. A biological exception might be the vanishing twin syndrome [11]. Unfortunately, there is always a small but important risk of technical variability due to contamination during blood sampling and centrifugation, plasma DNA isolation or during the PCR setup.

As there is no positive marker for the female sex, the absence of the Y-chromosomal sequences is interpreted as a predicted female fetus. A meta-analysis showed that fetal sex determination using cell-free fetal DNA has high sensitivity and specificity [12]. Nevertheless, even a low chance of error below 1% is of importance due to the use of this test for clinical decisions about the progress or discontinuation of pregnancies at sex-linked genetic risk. The test is prone to bias due to contamination with external male DNA. The SAFE collaborative project suggested multiple control mechanisms [13]. None of these widely accepted and followed suggestions is, however, able to distinguish contaminating external male DNA from fetal male DNA.

Using simple PCR targeted at Y-chromosomal sequences of various lengths we tested if it is possible to detect contaminating external male DNA in DNA samples from maternal plasma. We make use of the main biological feature of fetal DNA . its fragmentation [14]. The size of the fragments of fetal DNA present in maternal plasma varies but is mostly about 146 bp . the length the DNA from one nucleosome [15]. Others have shown that the maximum fragment length of cell-free fetal DNA circulating in maternal plasma comprises of 2 nucleosomes or less, but not exceeding 3 [16]. We suggest that the short length of fetal DNA fragments can be used not only for distinguishing fetal and maternal DNA [17], but also for distinguishing male fetal DNA and male contaminating (non-fetal) DNA. Quantitative real time PCR targeted at short and long Y-chromosomal sequences can detect male fetal DNA (positive only for short PCR targets) and male contaminating DNA (positive for both, short and long PCR fragments). In addition to this new technology based on novel PCR assays we suggest an algorithm for the use in a routine clinical laboratory focusing on not only in fetal sex determination.

**Materials and Methods**

*Samples*

Samples of peripheral blood taken from 60 pregnant women bearing female and male fetuses (in 10th . 20th weeks of gestation) were collected after informed consent was acquired. Blood samples were stored at 4 . 8 ° C and processed within 24 h. The plasma was separated by centrifugation two times at 2200 g for 10 min, then aliquoted to 1000 L and stored at . 20°C until further processing. Samples with lower than 1000 L volume of plasma were pooled in a way that it was possible to prepare 40 different pooled samples represented by 10 samples of female and 30 samples of male fetal gender. The pooled samples were prepared from one to three individual samples. Additionally, whole blood samples of ten males were used as control male samples (non-fetal) in the study.

*DNA extraction*

Circulating DNA (cDNA) was extracted from 40 individual or pooled plasma samples with use of DSP Virus Kit (Qiagen, Hilden, Germany). In the standard vacuum extraction protocol one modification was done, as the volume of plasma sample that should be processed is in original protocol 500 L and we have to start with 1000 L the volumes of protease, lysis buffer and ethanol were doubled in the first part of the DNA extraction protocol and the lysis was performed in 5 mL eppendorf tubes (Eppendorf, Hamburg, Germany). The isolated circulating DNA was eluted in 100 L of MBG water. In control male samples DNA was extracted from 200 L of whole blood with use of standard protocol and DNA Blood Mini Kit (Qiagen, Hilden, Germany).

### DNA sample utilization in the study

Of the isolated DNA samples . 40 cDNA and 10 control male samples, 20 of cDNA and 10 of control male samples were used in the first phase to test the ability of assays in qPCR and in detection of difference in performance of different assays regarding to the length of their amplicons. In the second phase 10 of remaining 20 cDNA samples (5:5 of male:female bearing pregnancies) were contaminated with control male DNA, while the other 10 (5:5 of male:female bearing pregnancies) were not contaminated. Samples were blinded, analysed and the statistical analysis of results of qPCR was performed.

### Contamination of extracted DNA samples

The concentration of male DNA used for contamination of isolated circulating DNA of pregnant women was determined with use of spectrophotometer and the DNA was diluted to the concentration of 50 pg/ L (~7.5 GEq/ L). Ten  L of the diluted contaminaning male DNA was added to the 65  L of isolated circulating DNA of pregnant women.

### Quantitative PCR

Isolated circulating DNA was quantified using real-time PCR with singleplex TaqMan probe assays. Four assays (one previously known, three novel assays) with lengths of amplicons between 84 bp and 290 bp targeted at *DYS14* sequence were designed with use of online tool Primer3 [http://biotools.umassmed.edu/bioapps/primer3_www.cgi]. The four assays were based on combination of 4 primers and one common TaqMan probe with sequences and corresponding amplicon lengths summarized in Table 1 and Figure 1, respectively.

Table1: Sequences of primers and probe used in amplification of four tested assays.

| Primer name | Sequence |
| --- | --- |
| DYS14-F | 5´-GGGCCAATGTTGTATCCTTCTC-3´ [10] |
| DYS14-R | 5´-GCCCATCGGTCACTTACACTTC-3´ [10] |
| DYS14-F_long | 5´-GTTAATGCCCAAGCCAGGA-3´ |
| DYS14-R_long | 5´-CAATAGTACCCACGCCTGCT-3´ |
| DYS14-probe | 5´- TCTAGTGGAGAGGTGCTC -3´ [10] |

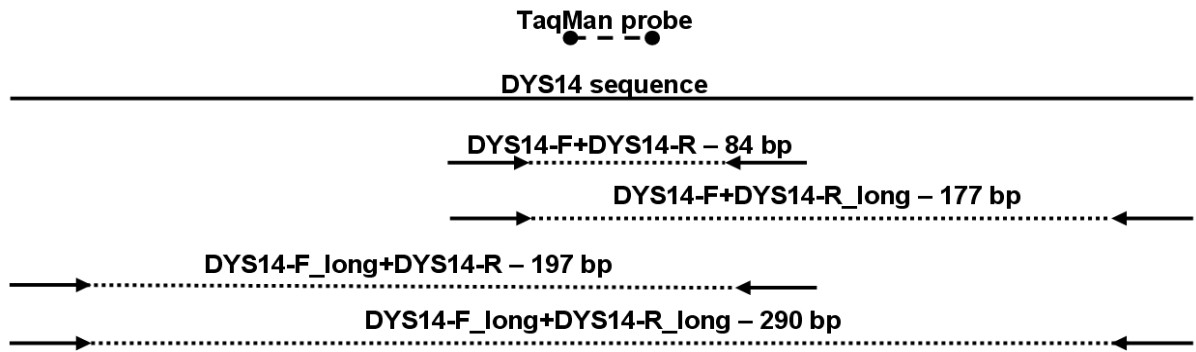

Figure1: Scheme of four DYS14 targeted amplicons tested in the study with their corresponding sizes.

PCR amplification was performed according to our previously published study [18]. Briefly, 5 μL of template DNA was used in 15 μL reaction volume containing final concentrations of 1 × QuantiFast Probe PCR Kit (Qiagen, Hilden, Germany), 1 μmol/L of each primer and 0.8 μmol/L of TaqMan probe. The following PCR program was used: initial denaturation step at 95 °C for 3 min, followed by 50 cycles each consisting of denaturation step at 95 °C for 3 s and combined annealing/extension step at 60 °C for 30 s. QPCR reactions run in the first phase in duplicates and in the second phase in triplicates. The real-time PCR was performed on Eppendorf Mastercycler ep realplex 4S (Eppendorf, Hamburg, Germany) and Roche LightCycler 480 (Roche, Rotkreuz, Switzerland) instruments with their corresponding softwares. The Ct value of 38 was set as the cuttof for positive signal of amplification.

### *Statistical analysis*

Data were analyzed using Microsoft Excel 2007, GraphPadPrism® version 5.03 and IBM SPSS 21. Blinded data were clustered using the K-Means clustering routine. Differences between grouped Ct values for the assay with the longer amplicon (177 bp) and the shortest amplicon (84 bp) were analyzed using One-way ANOVA with Bonferroni corrected post hoc t-test.

**Results**

### *Assays performance and fetal DNA amplification vs. control male DNA amplification differencies*

All 4 tested assays were found to be suitable to amplify their genomic targets with comparable qPCR efficiency (100.29% - 105.58%). When qPCR curve profiles were compared assays with shorter amplicons (84 bp and 177 bp) gave different profiles in comparison to the assays with longer amplicons (194 bp and 290 bp) (Figure 2). This was the reason that only the two shorter assays (84 bp and 177 bp) were used in subsequent analyses.

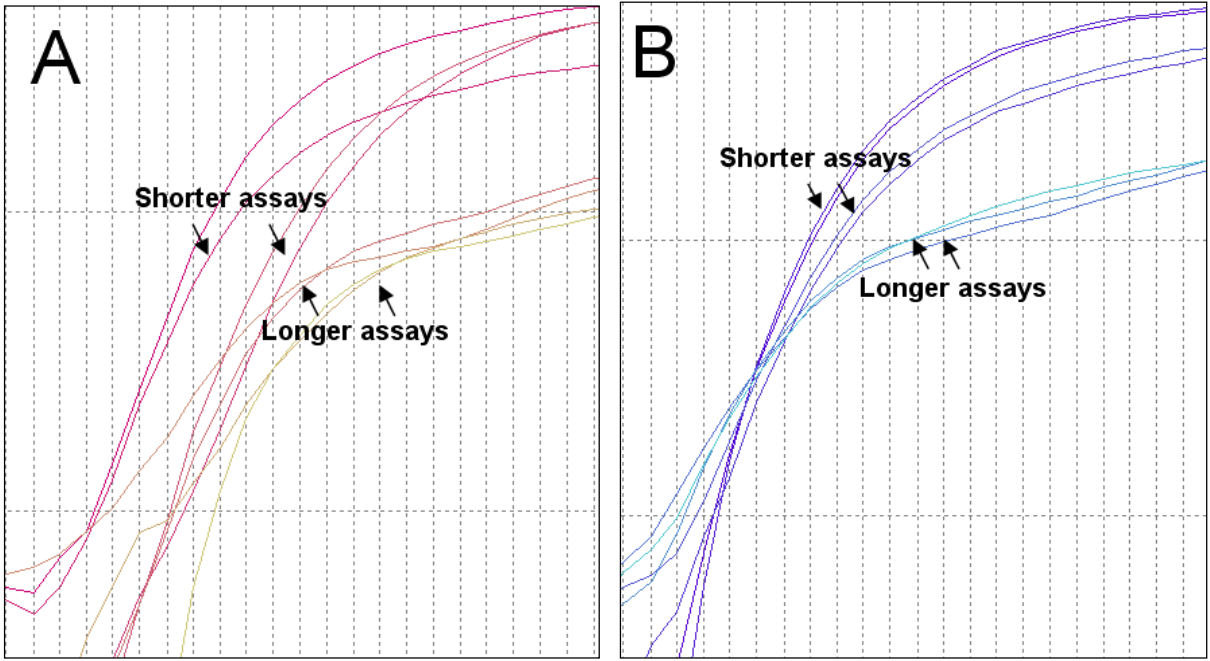

Figure 2: The comparison of amplification of DNA sample of pregnant woman bearing male fetus (A) and control male DNA (B) with all four tested assays represented with qPCR curve profiles reactions run in duplicates). Among shorter assays the 84 bp assay is the first from the left and the 177 bp assay the second from the left.

The difference in performance of the two tested assays with 84 bp and 177 bp amplicons between cDNA containing and control male DNA containing samples was recorded as the shift of Ct value in cDNA samples, while it was not present in control male DNA samples. In analysis of 20 cDNA samples of women bearing male fetuses and 10 control male DNA samples the difference measured as difference between Ct values of 177 bp and 84 bp assays was found to be statistically significant (p<0.0001, Figure 3).

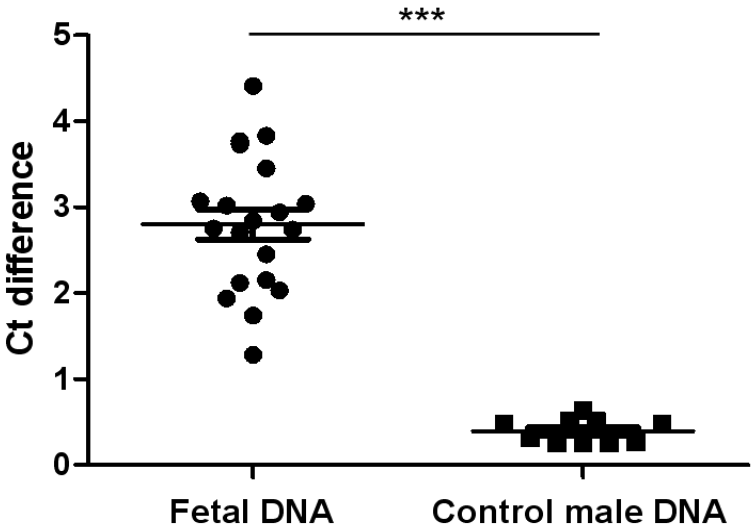

Figure 3: Differences between Ct values of 177 bp and 84 bp assays calculated for fetal and control male DNA samples (*** - p<0.0001).

## Analysis of contaminated vs. non-contaminated samples

The cluster analysis of blinded artificially contaminated samples correctly identified all contaminated samples as well as differentiated non-contaminated samples of pregnant women bearing male and female fetuses (F=34.4, p<0.0001, Figure 4).

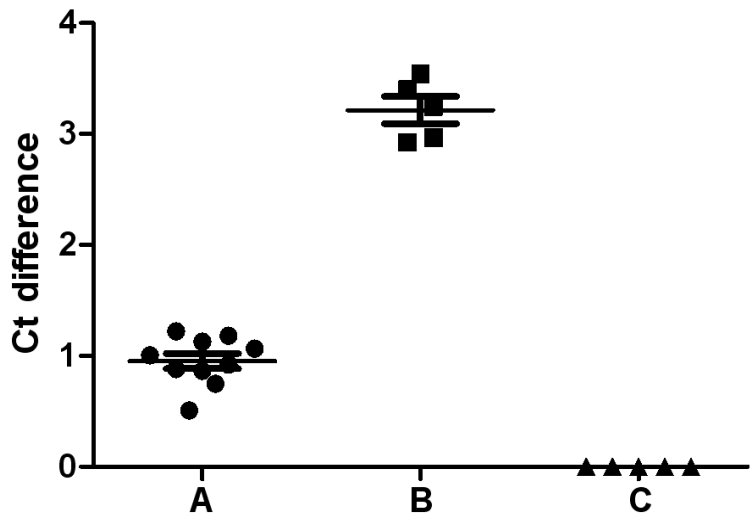

Figure 4: Group analysis and subsequent ANOVA analysis of contaminated and non-contaminated samples presented as differencies between Ct values of 177 bp and 84 bp assays (A - contaminated samples, B - non-contaminated samples of pregnant women bearing male fetuses, C - non-contaminated samples of pregnant women bearing female fetuses).

## Discussion

Although noninvasive prenatal gender testing made huge steps forward from the discovery of the presence of fetal DNA in maternal circulation in 1997, there were historically at least two issues which were needed to be solved before this test could be presented as truly clinical diagnostic test. These two issues were determination of presence of fetal DNA in analyzed sample in relevant amount (exclusion of false negatives) and detection of contamination by external male DNA during the sample processing (exclusion of false positives). The determination of presence of fetal DNA in relevant amount, which is sufficient for fetal gender testing, was recently successfully solved with the combination of utilization of differential methylation-based assay used previously known target *RASSF1A* with ultrasensitive method based on pyrophosphorolysis-activated polymerization [19]. In our study the novel approach based on known characteristics of fetal DNA in maternal circulation, the high level of fragmentation, was used in testing of possibility to detect contaminating male DNA. Usually in laboratories where non-invasive prenatal fetal gender tests are performed only females are working to minimize the possibility of test contamination with external male DNA. Such laboratory strategy is easily applicable to many laboratories, but the problem could arise when different than noninvasive prenatal fetal gender test should be performed, e.g.

noninvasive prenatal RhD test. As could be expected the contaminating external male DNA should differ from fetal DNA gained from maternal plasma in level of fragmentation. The external male DNA should be less fragmented than the fetal DNA, therefore qPCR assays targeted on the same chromosomal region but producing amplicons with specifically different sizes could be used for differentiation of contaminated and non-contaminated samples. To test this we utilize one known and designed 3 novel assays with amplicon sizes reaching 84 bp, 177 bp, 194 bp and 290 bp, all targeted to *DYS14* sequence. The assay with shortest amplicon (84 bp) was firstly described in 2005 [10] and afterward used in many laboratories for prenatal fetal gender determination. The other two assays (177 bp and 194 bp) were designed to be larger than the peak of cell free fetal DNA isolated from maternal plasma . 146 bp [15] and the last assay with largest amplicon (290 bp) should be close to the maximal DNA fragment size of cell free fetal DNA [16, 20]. In concordance with this knowledge we were able to get amplification from all four assays in cell free fetal DNA containing samples. All four assays performed well on control male DNA samples were these reached similar qPCR amplification efficiencies. In the first phase of the study, when only control male DNA and cDNA of pregnant women with male fetuses were analyzed, the expected shift in Ct values between shortest assay (84 bp amplicon) and all of the larger assays (177 bp, 194 bp and 290 bp amplicons) was recorded in cDNA samples, while it was not present in control male DNA samples (the lowest tested DNA amount in control male DNA samples was 5 genomic equivalents in qPCR reaction). But as the two shorter assays (84 bp and 177 bp) and two larger assays (194 bp and 290 bp) differ when amplification curves were compared and in some cell free fetal DNA samples the amplification in larger assays (194 bp and 290 bp) was not detected we decided not to continue with the larger assays in the second phase of the study as well to perform statistical analysis only for the two shorter assays (84 bp and 177 bp). The shift in Ct values was quantified as the difference of Ct value recorded for the larger assay (177 bp) and Ct value recorded for the shortest assay (84 bp). Statistical analysis with use of t-test reveals that the Ct difference between the tested groups was statistically significant ($p<0.0001$). With the calculated average shift (2.79, SD=±0.79) it was possible to estimate the proportion of DNA targets amplifiable with the shortest (84 bp) and the longer (177 bp) assay. There was on average 8-times (8.03, SD=±4.62) more DNA targets which were amplified by the shortest assay, what is in concordance with previous findings, where the proportion of cffDNA fragments in maternal plasma >193 bp was 20 % [17]. In the second phase of the project half (10) of the tested samples was artificially contaminated with control male DNA and blinded. The proportion of contaminating control male DNA to cffDNA was approx. 1:1 in terms of GEq in DNA solution used in qPCR amplification. In both groups of samples (contaminated and non-contaminated) half of the samples (5:5) were DNA from male bearing pregnancies and half from female fetus bearing pregnancies. After cluster analysis of Ct differences between the longer (177 bp) and the shorter (84 bp) assay three groups of samples were correctly identified, with group 1 containing 10 samples, which were artificially contaminated, and groups 2 and 3 (both with 5 samples) which contained non-contaminated samples from male and female fetus bearing pregnancies, respectively. ANOVA analysis performed after cluster analysis with the identified Ct differences in the three groups confirmed the results of the first phase.

**Conclusion**

Our study reveals the possibility to detect external contamination by male DNA in noninvasive prenatal fetal gender test with assays amplifying amplicons with different but specific sizes in qPCR. We designed simple, fast and cheap algorithm, which uses same target sequence for detection of true positive and false positive samples. For the detection of contamination the difference between Ct values of longer assay (177 bp) and shorter assay (84 bp) led to statistically significant results and clear differentiation of fetal and non-fetal male DNA samples. Moreover, the identified algorithm was successfully utilized in analysis of blinded cell-free fetal DNA containing samples which were artificially contaminated with external male DNA and correct clustering of all groups of samples was recorded. In the future our algorithm should be tested on different targets, e.g. RhD, what could confirm its universal applicability in assays used in noninvasive prenatal testing. This could lead to minimizing the proportion of inconclusive results, when external contamination could be the problem.

**Acknowledgements**

This publication is the result of implementation of the project: "Creating Competition Centre for research and development in the field of molecular medicine" (ITMS 26240220071) supported by the Research & Development Operational Programme funded by the ERDF."

**Conflict of interest**

The method described in this article is a subject-matter of the International Patent Application No. PCT/EP2014/068817 (unpublished, international filing date 04.09.2014, GENETON s.r.o., applicant).